\documentclass[a4paper,aps,prd,10pt,preprintnumbers,showpacs,twocolumn,superscriptaddress,nofootinbib,amsmath,amssymb]{revtex4-1}
\usepackage{cmap,graphicx,orcidlink}
\usepackage[utf8]{inputenc}
\usepackage[T1]{fontenc}

\def\imo{i}
\def\K{{\cal K}}
\def\re#1{Re(#1)}
\def\im#1{Im(#1)}

\def\Order#1{{\cal O}\left(#1\right)}

\begin{document}

\title{Correspondence between grey-body factors and quasinormal frequencies for rotating black holes}

\author{R. A. Konoplya \orcidlink{0000-0003-1343-9584}}
\email{roman.konoplya@gmail.com}
\affiliation{Research Centre for Theoretical Physics and Astrophysics, Institute of Physics, Silesian University in Opava, Bezručovo náměstí 13, CZ-74601 Opava, Czech Republic}

\author{A. Zhidenko \orcidlink{0000-0001-6838-3309}}
\email{olexandr.zhydenko@ufabc.edu.br}
\affiliation{Centro de Matemática, Computação e Cognição (CMCC), Universidade Federal do ABC (UFABC), \\ Rua Abolição, CEP: 09210-180, Santo André, SP, Brazil}

\begin{abstract}
Although the proper oscillation frequencies of black holes (quasinormal modes) and the grey-body factors, which determine the scattering properties of black holes, represent two distinct spectral problems with different boundary conditions, a recent study has revealed an intrinsic connection between these quantities. We have shown that the correspondence between grey-body factors and quasinormal modes, previously established for spherically symmetric and asymptotically flat black holes, also extends to general parametrized axially symmetric black holes. This correspondence is limited to non-superradiant waves.
\end{abstract}

\maketitle

\section{Introduction}

In black-hole physics, grey-body factors are essential for understanding the spectrum of radiation that escapes from a black hole. These factors account for the deviation from perfect blackbody radiation due to the scattering and absorption effects caused by the black hole's gravitational potential. Unlike the idealized blackbody scenario, grey-body factors provide a more realistic description of how various fields -— such as electromagnetic, or gravitational waves —- are transmitted through the black hole's potential barrier.

Quasinormal modes are the characteristic oscillations of a black hole, defined by complex frequencies that satisfy specific boundary conditions: purely outgoing waves at infinity and purely ingoing waves at the event horizon.
Grey-body factors are the transmission coefficients that describe the fraction of radiation emitted by a black hole that can escape to infinity. They are calculated by solving the wave equation with boundary conditions of purely ingoing waves at the event horizon and a combination of incoming and outgoing waves at infinity. Despite different boundary conditions and roles of these two characteristics, there are links between them, as has been recently discussed in \cite{Konoplya:2017wot,Oshita:2022pkc,Oshita:2023cjz,Rosato:2024arw,Oshita:2024fzf,Oshita:2024wgt,Konoplya:2024lir,Albuquerque:2024cwl,Pedrotti:2024znu}. In particular, for spherically symmetric amd asymptotically flat black holes a correspondence has been established connecting the fundamental quasinormal mode and grey-body factors in the eikonal limit \cite{Konoplya:2024lir}. This correspondence is valid as along as the WKB approach describes the eikonal regime adequately, in a similar way with the correspondence between between null geodesics and eikonal quasinormal modes \cite{Cardoso:2008bp,Khanna:2016yow,Konoplya:2017wot,Konoplya:2022gjp,Bolokhov:2023dxq,Konoplya:2020bxa}. Recently the validity of the correspondence between grey-body factors and quasinormal modes has been tested and used for a number of spherically symmetric spacetimes \cite{Skvortsova:2024msa,Dubinsky:2024vbn,Malik:2024cgb} and extended to spherically symmetric wormholes \cite{Bolokhov:2024otn}.

Here, we aim to take the next step and explore whether the correspondence can be extended to the axially symmetric case. While in the spherically symmetric scenario, the correspondence can be derived under relatively general assumptions without specifying the black-hole metric, the axially symmetric case is more complex. This complexity arises because the separation of variables is necessary to verify the correspondence. Consequently, we begin with the general Konoplya-Rezzolla-Zhidenko (KRZ) parameterization for axially symmetric black holes \cite{Rezzolla:2014mua,Konoplya:2016jvv,Younsi:2016azx} and limit our analysis to spacetimes that allow for the separation of variables, as described in \cite{Konoplya:2018arm}.

We will show that the expression for qrey-body factors in terms of
quasinormal modes which was derived for spherically symmetric black holes in \cite{Konoplya:2024lir} is valid also for axially symmetric black holes without modification.

The paper is organized as follows. In Sec.~\ref{sec:basicequations}, we review the axially symmetric and asymptotically flat black hole spacetime that allows for the separation of variables in the Klein-Gordon equation and the resultant wave-like equation. Sec.~\ref{sec:numerical} is dedicated to the accurate numerical methods for calculating the quantities on both sides of the correspondence: grey-body factors and quasinormal modes. In Sec.~\ref{sec:res}, we compute these quantities and compare them to demonstrate that the relationship derived earlier for the spherically symmetric case also holds for axial symmetry. Finally, in the Conclusion, we summarize the results obtained and discuss some open questions.

\section{Parametrized black hole metric and wave equation}\label{sec:basicequations}

The KRZ parametrization \cite{Konoplya:2016jvv} offers a general framework for describing axially symmetric and asymptotically flat black holes. The KRZ parametrization provides a model-independent and systematic method for studying deviations from the Kerr metric, which is particularly valuable for testing theories of gravity beyond general relativity using observational data, such as those from gravitational wave detectors or X-ray observations of black hole accretion disks. Similar in spirit to the post-Newtonian parametrized formalism, the KRZ approach is distinct in its validity across the entire space outside the black hole.

This approach is based on a double expansion: one in the radial direction, represented as a continued fraction in terms of a compact coordinate, and another around the equatorial plane as a series of $\cos \theta$. The continued fraction expansion offers superior convergence, and as demonstrated in previous studies, only a few coefficients are typically needed to accurately describe the black hole. Given the extensive application and discussion of this method in numerous works \cite{Kocherlakota:2020kyu,Zhang:2024rvk,Cassing:2023bpt,Li:2021mnx,Ma:2024kbu,Bronnikov:2021liv,Shashank:2021giy,Konoplya:2021slg,Kokkotas:2017zwt,Yu:2021xen,Toshmatov:2023anz,Konoplya:2019fpy,Nampalliwar:2019iti,Ni:2016uik,Konoplya:2018arm,Zinhailo:2018ska,Paul:2023eep}, we will summarize only its main points here.

General axially symmetric black hole metric depends on the five metric functions. The line element can be written in the following general way \cite{Konoplya:2016jvv}:
\begin{eqnarray}
ds^2 &=&
-\dfrac{N^2(r,\theta)-W^2(r,\theta)\sin^2\theta}{K^2(r,\theta)}dt^2
\nonumber \\&&
-2W(r,\theta)r\sin^2\theta dt \, d\phi
+K^2(r,\theta)r^2\sin^2\theta d\phi^2
\nonumber \\&&
+\Sigma(r,\theta)\left(\dfrac{B^2(r,\theta)}{N^2(r,\theta)}dr^2 +
r^2d\theta^2\right). \label{eq:initmetric}
\end{eqnarray}

We will further restrict our analysis to metrics where the variables in the general covariant Klein-Gordon equation can be separated, as outlined in \cite{Konoplya:2018arm}. In this case, there are three independent metric functions,
$R_{N}(r)$, $R_{B}(r)$, $R_{\Sigma}(r)$, which are related to the previously mentioned metric functions as follows:
\begin{eqnarray}
\nonumber
N^2(r,\theta)&=&R_N(r),\qquad\qquad B(r,\theta)=R_B(r),\\
\label{eq:gen}
\Sigma(r,\theta)&=&R_\Sigma(r)+\dfrac{a^2\cos^2\theta}{r^2},\\
\nonumber
W(r,\theta)&=&\dfrac{a}{r\Sigma(r,\theta)}\left(R_\Sigma(r)+\frac{a^2}{r^2}-R_N(r)\right),\\
\nonumber
K^2(r,\theta)&=&\dfrac{1}{\Sigma(r,\theta)}\left(\left(R_\Sigma(r)+\dfrac{a^2}{r^2}\right)^2\!\!\!\!-\dfrac{a^2\sin^2\theta}{r^2}R_N(r)\right).
\end{eqnarray}
Here $a$ is the rotation parameter.
The massless scalar field obeys the following general covariant Klein-Gordon equation
\begin{equation}
\Box\Phi(t,r,\theta,\phi)= \frac{1}{\sqrt{-g}}\partial_\mu \left(\sqrt{-g}g^{\mu \nu}\partial_\nu\Phi(t,r,\theta,\phi) \right)=0.
\end{equation}
In the background given by the metric (\ref{eq:gen}) after using the following ansatz
\begin{equation}\label{eq:ansatz}
\Phi(t,r,\theta,\phi)=e^{-\imo\omega t+\imo m\phi}\Psi(r)S(\cos\theta),
\end{equation}
the Klein-Gordon equation is reduced to the following master equation for the radial function $\Psi(r)$,
\begin{eqnarray}\label{eq:master}
&&\Biggr(\dfrac{R_N(r)}{R_B(r)}\dfrac{d}{dr}\dfrac{r^2 R_N(r)}{R_B(r)}\dfrac{d}{dr}+\dfrac{(r^2\omega R_\Sigma(r) + a^2\omega-a m)^2}{r^2}\nonumber\\&&-R_N(r)\lambda(\omega)\Biggr)\Psi(r)=0.
\end{eqnarray}

Using the new variable $y\equiv\cos\theta$, the eigenvalues $\lambda(\omega)$ can be found by solving eigenvalue problem for the spheroidal wave equation,
\begin{eqnarray}
&&\Biggr(\dfrac{d}{dy}(1-y^2)\dfrac{d}{dy}+\omega^2a^2y^2-\dfrac{m^2y^2}{1-y^2}\\\nonumber&&-(m-\omega a)^2+\lambda(\omega)\Biggr)S(y)=0,
\end{eqnarray}
The above equation can be solved numerically and the eigenvalues can be found using the standard Wolfram Mathematica\textregistered{} built-in function
$$\lambda(\omega)=\mathrm{SpheroidalEigenvalue}(\ell,m,\imo \omega a)-2m\omega a$$
for a given integer $\ell=|m|,|m|+1,|m|+2\ldots$.

Following \cite{Rezzolla:2014mua,Konoplya:2016jvv}, we define the event horizon radius $r_0$ as the largest solution to the equation
$$R_N(r)=0,$$
and introduce the compact coordinate
$$x\equiv1-\frac{r_0}{r}.$$
Then the metric functions are defined via the infinite continued fractions,
\begin{eqnarray}
R_\Sigma&=&1\\
R_B&=&b_{00}(1-x)+\dfrac{b_{01}(1-x)^2}{1+\dfrac{b_{02}x}{1+\dfrac{b_{03}x}{1+\ldots}}}\,,\\
R_N&=&x\Biggr(1-(1-x)\epsilon_0
\\\nonumber&&-(1-x)\left(a_{00}-\epsilon_0+\frac{a^2}{r_0^2}\right)+\dfrac{a_{01}(1-x)^2}{1+\dfrac{a_{02}x}{1+\dfrac{a_{03}x}{1+\ldots}}}\Biggr)\,.
\end{eqnarray}
Here the coefficient $\epsilon_0$ relates the event horizon radius $r_0$ and the asymptotic mass $M$,
\begin{equation}
\epsilon_0=\frac{2 M-r_0}{r_0}.
\end{equation}
The coefficients $a_{00}$ and $b_{00}$ depend on the post-Newtonian parameters and, according to the current constrains in the weak field regime, must be small, so that in the present paper we assume that
$$a_{00}=b_{00}=0.$$

The coefficients $a_{01},a_{02},a_{03},\ldots$ and $b_{01},b_{02},b_{03},\ldots$ are responsible for the near-horizon deviations of the geometry from the Kerr limit. Since for the Kerr black hole $\epsilon_0=a^2/r_0^2$, we introduce also the additional deformation parameter $\delta_M$,
\begin{equation}
\epsilon_0=\frac{a^2}{r_0^2}+\delta_M,
\end{equation}
which corresponds to the deviation of the black-hole mass from its Kerr value.

\section{Numerical methods}\label{sec:numerical}

\subsection{Accurate calculation of the quasinormal modes}

Quasinormal modes are the solutions to the eigenvalue problem for
$\omega$ defined by Eq.~(\ref{eq:master}). These solutions represent waves that are purely ingoing at the event horizon and purely outgoing at infinity. To accurately determine the quasinormal modes, we employ the Leaver method \cite{Leaver:1985ax}, which relies on a convergent process utilizing the Frobenius series expansion.
We introduce the new function
\begin{equation}\label{eq:reg}
\Psi(r)=e^{\imo\omega r}(r-r_1)^{\alpha}\left(\frac{r-r_0}{r-r_1}\right)^{-\imo(\omega-\Omega_m)/2\kappa}y(r),
\end{equation}
where $r_1$ denotes the inner horizon $r_1<r_0$, and
$$\Omega_m\equiv\frac{ma}{r_0^2+a^2}.$$
The constant $\alpha$ is defined in such a way that $y(r)$ is regular at $r=\infty$ once $\Psi(r)$ corresponds to the purely outgoing wave at spatial infinity, while $\kappa>0$ is chosen in order to have the regular $y(r)$ at $r=r_0$ once $\Psi(r)$ corresponds to the purely ingoing wave at the horizon.

We expand the regular function $y(r)$ using the following Frobenius series:
\begin{equation}\label{eq:Frobenius}
y(r)=\sum_{k=0}^{\infty}a_k\left(\frac{r-r_0}{r-r_1}\right)^k,
\end{equation}
and find that the coefficients $a_k$ satisfy the n-term recurrence relation, which can be reduced to the three-term recurrence relation via Gaussian elimination (see, for example, \cite{Konoplya:2011qq} for details). Then, using the coefficients in the recurrence relation, we find an infinite continued fraction equation with respect to $\omega$, which is satisfied when the series (\ref{eq:Frobenius}) converges at $r=\infty$, or, in other words, if $\Psi(r)$ satisfies the quasinormal boundary conditions. In order to obtain the final equation we employ a sequence of positive real midpoints as described in \cite{Rostworowski:2006bp} and use the Nollert improvement~\cite{Nollert:1993zz}, which was generalized in~\cite{Zhidenko:2006rs} for the recurrence relation of arbitrary number of terms.

\subsection{Grey-body factors}
For the scattering problem we consider the following boundary conditions
\begin{equation}
\begin{array}{rclcl}
\Psi &=& A_i \Psi_i(r) + A_o \Psi_o(r), &\quad& r\to\infty, \\
\Psi &=& A_t \Psi_t(r), &\quad& r\to r_0,
\end{array}
\end{equation}
where $\Psi_i(r)$ corresponds to the incident wave, $\Psi_t(r)$ is the transmitted wave falling to the black hole, and $\Psi_o(r)$ is the reflected outgoing wave at infinity.

The grey-body factor is defined as
\begin{equation}\label{eq:greybody}
\Gamma(\omega)=1-\left|\frac{A_o}{A_i}\right|^2.
\end{equation}

In order to calculate the grey-body factors precisely, for given value of real $\omega$, we numerically integrate Eq.~(\ref{eq:master}) using the Runge-Kutta method from the point $r_i>r_0$, where we impose the initial conditions obtained from the series expansion of the ingoing wave at the horizon
\begin{eqnarray}\nonumber
\Psi(r)&=&\left(r-r_0\right)^{-\imo(\omega-\Omega_m)/2\kappa}\Biggl(1+C_1\left(r-r_0\right)
\\\label{eq:horizon}
&&+C_2\left(r-r_0\right)^2+C_3\left(r-r_0\right)^3+\dots\Biggr).
\end{eqnarray}
In the above equations we use the sufficient number of coefficients $C_1,C_2,\ldots$, in order to obtain sufficiently accurate initial conditions at $r=r_i$.

Finally we fit the numerical solution for $r\gg r_0$ by the superposition of the expansions for the ingoing and outgoing waves,
\begin{eqnarray}
\Psi(r\gg r_0)&\approx& A_ie^{\imo\omega r}r^{\alpha}\left(1+\frac{F_1}{r}+\frac{F_2}{r^2}+\ldots\right)
\\\nonumber
&+&A_oe^{-\imo\omega r}r^{-\alpha}\left(1+\frac{G_1}{r}+\frac{G_2}{r^2}+\ldots\right),
\end{eqnarray}
where again we compute the sufficient number of the coefficients $F_1,F_2,\ldots$ and $G_1,G_2,\ldots$ from the asymptotic expansion of Eq.~(\ref{eq:master}) in order to obtain the accurate values for the ingoing and outgoing coefficients, $A_i$ and $A_o$, which are used to calculate the grey-body factor (\ref{eq:greybody}).

With the above numerical methods allowing us to find quasinormal modes and grey-body factors precisely, we are ready to study the correspondence between these quantities.

\section{Correspondence between the quasinormal modes and grey-body factors}\label{sec:res}

\begin{figure}
\resizebox{\linewidth}{!}{\includegraphics{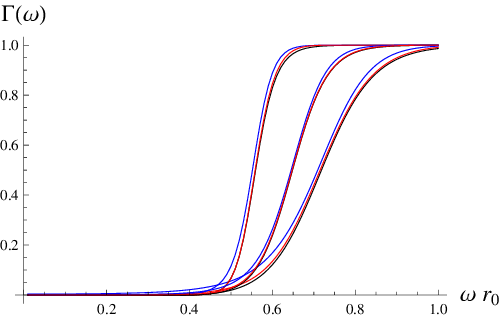}}
\caption{Accurate grey-body factors (black) and the approximations by the eikonal formula (blue) and the second-order formula (red) for $\ell=m=1$, $a=0.5r_0$, $a_1=b_1=0$. Left: $\delta_M=0.25$ ($\omega_0r_0=0.551768-0.076079\imo$, $\omega_1r_0=0.497689-0.260286\imo$), middle: $\delta_M=0$ -- Kerr black hole ($\omega_0r_0=0.645236-0.133012\imo$, $\omega_1r_0=0.625177-0.4030240\imo$), right: $\delta_M=-0.25$ ($\omega_0r_0=0.709286-0.198484\imo$, $\omega_1r_0=0.661855-0.611948\imo$).}\label{fig:l1m1a05dM}
\end{figure}

\begin{figure}
\resizebox{\linewidth}{!}{\includegraphics{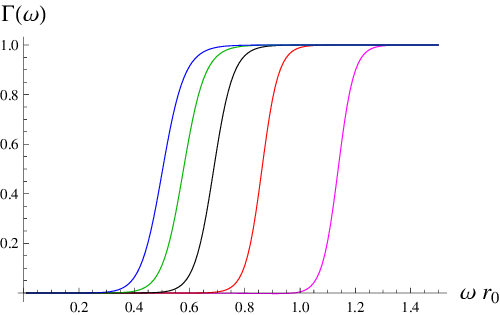}}
\resizebox{\linewidth}{!}{\includegraphics{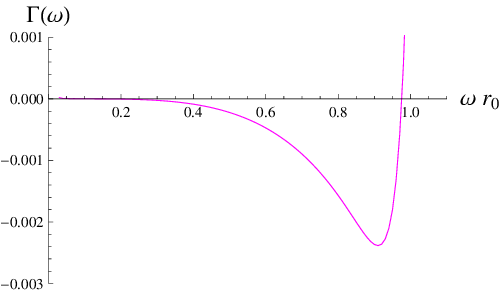}}
\caption{Upper panel: Accurate grey-body factors for the non-Kerr black hole $a=0.8r_0$, $\delta_M=-0.25$, $a_1=0.2$, $a_2=4$, $a_3=0$, $b_1=0.5$, $b_2=0$ for $\ell=2$ and (from left to right) $m=-2$ (blue), $m=-1$ (green), $m=0$ (black), $m=1$ (red), $m=2$ (magenta). Lower panel: negative factor due to superradiance for $m=2$.}\label{fig:l2a08nonKerr}
\end{figure}

\begin{figure}
\resizebox{\linewidth}{!}{\includegraphics{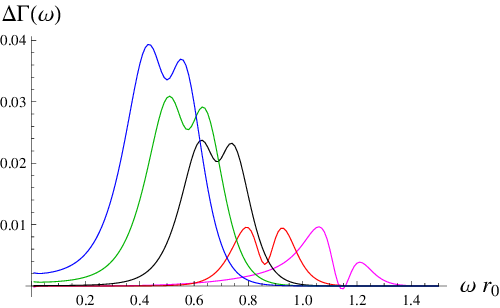}}
\resizebox{\linewidth}{!}{\includegraphics{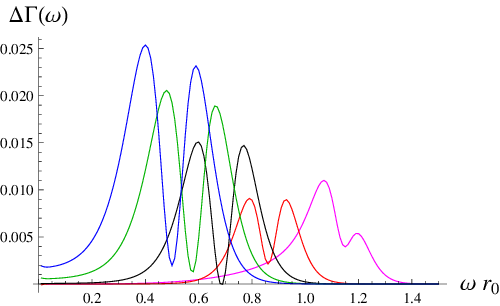}}
\resizebox{\linewidth}{!}{\includegraphics{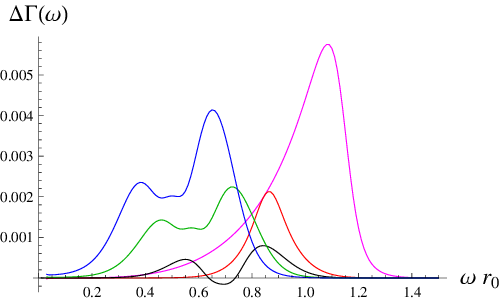}}
\caption{Difference between the approximate and the accurate values of the grey-body factors for the non-Kerr black hole $a=0.8r_0$, $\delta_M=-0.25$, $a_1=0.2$, $a_2=4$, $a_3=0$, $b_1=0.5$, $b_2=0$ for $\ell=2$ and (from left to right) $m=-2$, $m=-1$, $m=0$, $m=1$, $m=2$: eikonal formula (upper panel), first-order formula (middle panel), second-order formula (lower panel). The values of the fundamental mode and the first overtone, which are used in the approximate formulas, were calculated with the help of the continued fraction method.}\label{fig:l2a08nonKerrdiff}

\begin{tabular*}{\linewidth}{@{\extracolsep{\fill}}rccc}
\\
\hline
$m$&color&$\omega_0r_0$&$\omega_1r_0$ \\
\hline
$-2$ & blue    & $0.497498-0.125396\imo$ & $0.457102-0.388311\imo$ \\
$-1$ & green   & $0.574332-0.121536\imo$ & $0.544331-0.371840\imo$ \\
$~0$ & black   & $0.684671-0.111143\imo$ & $0.661595-0.334768\imo$ \\
$~1$ & red     & $0.861535-0.094007\imo$ & $0.860206-0.281146\imo$ \\
$~2$ & magenta & $1.137820-0.090992\imo$ & $1.142070-0.275444\imo$ \\
\hline
\end{tabular*}
\end{figure}

Recently, an analytic formula was derived in \cite{Konoplya:2024lir} that enables the calculation of grey-body factors, given that the fundamental mode and the first overtone are known. If the effective potential belongs to a class that can be addressed using the WKB approach \cite{Konoplya:2019hlu}, this formula is exact in the eikonal regime, i.e., as $\ell\to\infty$ and provides an approximation at smaller $\ell$. However, by including correction terms beyond the eikonal order, the formula can achieve sufficient accuracy even at moderate values of $\ell$.

The formulas were derived for the spherically symmetric and asymptotically flat black holes via the WKB expression for the grey-body factors,
\begin{equation}\label{eq:gbfactor}
\Gamma_{\ell}(\omega)=\dfrac{1}{1+e^{2\pi\imo \K}},
\end{equation}
where $\K$ satisfies the equation
\begin{eqnarray}\label{WKBformula-spherical}
\omega^2&=&V_0+A_2(\K^2)+A_4(\K^2)+A_6(\K^2)+\ldots\\\nonumber
&-&\imo\K\sqrt{-2V_2}\left(1+A_3(\K^2)+A_5(\K^2)+A_7(\K^2)\ldots\right).
\end{eqnarray}
Here, $V_0$ is the value of the effective potential at its maximum, $V_2$ is the value of the second derivative of the potential at this point, and $A_i$ for $i=2, 3, 4, \ldots$ are the $i$-th order WKB correction terms beyond the eikonal approximation. These terms depend on $\K$ and the derivatives of the potential at its maximum up to the order $2i$. Explicit forms of $A_i$ are provided in \cite{Iyer:1986np} for the second and third WKB orders, in \cite{Konoplya:2003ii} for the fourth to sixth orders, and in \cite{Matyjasek:2017psv} for the seventh to thirteenth orders.

Relation~(\ref{WKBformula-spherical}) also allows for the calculation of the quasinormal modes. Due to the boundary conditions for quasinormal modes, the following matching arises:
\begin{equation}
\K=n+\frac{1}{2}, \quad n=0,1,2,\ldots.
\end{equation}
Here, $n$ is the overtone number: For $n = 0$ ($\K = 1/2$), formula~(\ref{WKBformula-spherical}) yields the dominant quasinormal frequency $\omega_0$, which corresponds to the mode with the slowest decay rate.

In~\cite{Konoplya:2024lir}, it was demonstrated that, by expanding formula~(\ref{WKBformula-spherical}) in the regime of large $\ell$, the grey-body factor can be expressed entirely in terms of the quasinormal modes. In this expansion, all derivatives of the potential at its maximum cancel out from the final expression for the grey-body factors. Therefore, when the WKB approximation is valid, the grey-body factors are determined solely by the values of the quasinormal modes.

In particular, within the eikonal approximation, we have
\begin{equation}\label{eq:gbeikonal}
\imo\K=\frac{\omega^2-\re{\omega_0}^2}{4\re{\omega_0}\im{\omega_0}}+\Order{\frac{1}{\ell}}.
\end{equation}

The first-order beyond eikonal correction is given by
\begin{eqnarray}\label{eq:gbfirstorder}
\imo\K=\dfrac{\omega^2 - \re{\omega_0}^2}{4\re{\omega_0}\im{\omega_0}}-\dfrac{\re{\omega_0}-\re{\omega_1}}{16\im{\omega_0}},
\end{eqnarray}
where we introduced the frequency of the first overtone $\omega_1$, which can be obtained from (\ref{WKBformula-spherical}) for $n=1$ ($\K=3/2$).

A more accurate formula including the second-order correction beyond the eikonal limit is
\begin{eqnarray}\nonumber
&&\imo\K=\frac{\omega^2-\re{\omega_0}^2}{4\re{\omega_0}\im{\omega_0}}\Biggl(1+\frac{(\re{\omega_0}-\re{\omega_1})^2}{32\im{\omega_0}^2}
\\\nonumber&&\qquad\qquad-\frac{3\im{\omega_0}-\im{\omega_1}}{24\im{\omega_0}}\Biggr)
-\frac{\re{\omega_0}-\re{\omega_1}}{16\im{\omega_0}}
\\\nonumber&& -\frac{(\omega^2-\re{\omega_0}^2)^2}{16\re{\omega_0}^3\im{\omega_0}}\left(1+\frac{\re{\omega_0}(\re{\omega_0}-\re{\omega_1})}{4\im{\omega_0}^2}\right)
\\\nonumber&& +\frac{(\omega^2-\re{\omega_0}^2)^3}{32\re{\omega_0}^5\im{\omega_0}}\Biggl(1+\frac{\re{\omega_0}(\re{\omega_0}-\re{\omega_1})}{4\im{\omega_0}^2}
\\\nonumber&&\qquad +\re{\omega_0}^2\Biggl(\frac{(\re{\omega_0}-\re{\omega_1})^2}{16\im{\omega_0}^4}
\\&&\qquad\qquad -\frac{3\im{\omega_0}-\im{\omega_1}}{12\im{\omega_0}}\Biggr)\Biggr)+ \Order{\frac{1}{\ell^3}}.
\label{eq:gbsecondorder}
\end{eqnarray}

Since the above formulas depend solely on the quasinormal modes and not explicitly on the multipole number $\ell$, one might conjecture that this approximate relationship between quasinormal modes and grey-body factors is valid for arbitrary, or at least a broad class of, black hole
spacetimes. This holds if the WKB approach is valid, i.e. if the turning points are sufficiently close.

From Fig.~\ref{fig:l1m1a05dM}, we observe that the eikonal formula provides a good approximation for the grey-body factors for both Kerr and non-Kerr black holes. The second-order correction further improves accuracy, reducing the difference between the approximate formula and the exact grey-body factor values to less than $0.015$.

While the second-order expansion in terms of $1/\ell$ incorporates only the fundamental mode and the first overtone, we anticipate that higher-order expansions may include higher overtones as well. However, deriving higher-order expressions that link grey-body factors with quasinormal modes is intricate and would necessitate a dedicated study, along with a rigorous analysis of the contributions from higher-order terms.

It is important to note that the derived formula does not account for superradiance, as it inherently yields only positive values for the grey-body factors. Consequently, this correspondence cannot be used to analyze superradiance.

In order to test the correspondence we compare the {grey-body factors obtained via the approximate formula (\ref{eq:gbsecondorder}), where the quasinormal modes are found precisely with the help of the Frobenius method with the precise values of the grey-body factors found by numerically integrating the wave-like equation (\ref{eq:master}). For illustrations we choose rotating Kerr and non-Kerr black holes (see Fig.~\ref{fig:l1m1a05dM}).

We also consider the grey-body factors for various values of the multipole number for the rapidly rotating non-Kerr black hole, whose geometry significantly differs from the Kerr one with the same rotation parameter (see Fig.~\ref{fig:l2a08nonKerr}).
From Fig.~\ref{fig:l2a08nonKerrdiff} we observe that:

\begin{enumerate}
  \item The approximate formulas can be applied to the rapidly rotating black holes and provide a good approximation.
  \item The eikonal formula works generally better for positive $m$, when $|\re{\omega_0}| \gg |\im{\omega_0}|$.
  \item The formula including beyond eikonal corrections provides improved accuracy, although the best approximation is obtained for $m=0$.
  \item For $\ell=m=2$ the convergence is the slowest, partially due to the significant contribution from the superradiance, which cannot be reproduced by the correspondence.
\end{enumerate}

In the eikonal regime, where $\ell\to\infty$, the relation (\ref{eq:gbeikonal}) becomes exact for the considered class of axially symmetric black holes. This can be observed by choosing a large $\ell \gg 1$ and comparing the grey-body factor obtained via (\ref{eq:gbeikonal}) with the precise values.

\section{Conclusions}

The relation between grey-body factors and quasinormal modes was recently derived for spherically symmetric and asymptotically flat black holes \cite{Konoplya:2024lir}. Considering recent observations that, unlike quasinormal modes, grey-body factors are significantly more stable to small deformations of the spacetime geometry —- such as those caused by astrophysical environments—grey-body factors \cite{Rosato:2024arw,Oshita:2024fzf} may become a useful characteristic not only in studying Hawking radiation but also in classical radiation phenomena.

However, whether this relation between grey-body factors and quasinormal modes can be applied to realistic black hole configurations depends on whether it can be extended to the case of rotating, and even rapidly rotating, black holes. This extension is constrained by the separability of variables in the perturbation equations. For illustration, we consider the case where variables can be separated in the Klein-Gordon equation. Our systematic approach to the black hole spacetimes under consideration is guided by the general KRZ parametrization for axially symmetric and asymptotically flat black holes in arbitrary metric theories of gravity \cite{Konoplya:2016jvv}. Several examples used for illustration demonstrate that the relation connecting grey-body factors and quasinormal modes, both in the eikonal regime and beyond, also holds for rotating black holes.

Nevertheless, this relation does not account for the effect of superradiance, as corrections beyond the eikonal regime —- though providing a better approximation than the eikonal formula —- do not guarantee quick convergence to the precise result. This limitation is related to the WKB origin of the correspondence: the WKB series converges only asymptotically and does not guarantee convergence at every subsequent order. This raises the main open question regarding the observed link between grey-body factors and quasinormal modes: Is there a more general correspondence, beyond and independent of the WKB arguments we used? Regardless of the answer, the discovered relation can serve as a reliable approximation in many cases.

While the employed form of the WKB formula assumes that the black hole is either asymptotically flat or de Sitter, it is intriguing to explore whether a similar correspondence between quasinormal modes and grey-body factors exists for asymptotically anti-de Sitter black holes and branes. In such cases, quasinormal modes provide a means to describe conformal thermal field theory at strong coupling \cite{Kovtun:2004de,Son:2007vk}. This correspondence could, in particular, be applied to describe the dispersion relations and hydrodynamic regime of strongly coupled systems \cite{Son:2007vk,Dodelson:2023vrw}, where the thermal two-point correlator is also fixed by the quasinormal modes.

\bibliography{bibliography}

\end{document}